\documentclass[aps,twocolumn,showpacs,preprintnumbers,prl]{revtex4}

\usepackage{graphicx}
\usepackage{dcolumn}
\usepackage{bm}

\begin{document}


\title{The role of surface plasmons in the decay of image-potential
states\\ on silver surfaces}
\author{A. Garc\'\i a-Lekue$^{1}$, J. M. Pitarke$^{1,2}$, E. V.
Chulkov$^{2,3}$, A. Liebsch$^{4}$, and P. M. Echenique$^{2,3}$}
\affiliation{$^1$ Materia Kondentsatuaren Fisika Saila, Zientzi Fakultatea,
Euskal Herriko Unibertsitatea,\\
644 Posta kutxatila, E-48080 Bilbo, Basque Country\\
$^2$Donostia International Physics Center (DIPC) and Centro Mixto
CSIC-UPV/EHU,\\ Manuel de Lardizabal Pasealekua, E-20018 Donostia, Basque
Country\\
$^3$Materialen Fisika Saila, Kimika Fakultatea, Euskal Herriko
Unibertsitatea,\\ 
1072 Posta kutxatila, E-20080 Donostia, Basque Country\\
$^4$Institut f\"ur Festk\"orperforschung, Forschungszentrum J\"ulich,
52425 J\"ulich, Germany}                        
\date{\today}

\begin{abstract}
The combined effect of single-particle and collective surface excitations
in the decay of image-potential states on Ag surfaces is investigated, and
the origin of the long-standing discrepancy between experimental measurements
and previous theoretical predictions for the lifetime of these states is
elucidated. Although surface-plasmon excitation had been expected to reduce
the image-state lifetime, we demonstrate that the subtle combination of the
spatial variation of $s$-$d$ polarization in Ag and the characteristic
non-locality of many-electron interactions near the surface yields
surprisingly long image-state lifetimes, in agreement with experiment.  
\end{abstract}

\pacs{71.45.Gm, 73.20.At, 73.50.Gr, 78.47.+p}

\maketitle

The understanding of charge localization and electron
dynamics at solid surfaces is of crucial importance in a large variety of
physical and chemical phenomena, such as energy transfer, electronically
induced adsorbate reactions, catalytic processes, epitaxy, and the development
of new materials \cite{science1,science2,science3,nienhaus}. Accordingly, over
the last two decades there has been a large effort to investigate the
characteristic energies and linewidths of short-lived electronic excitations
at metal surfaces  \cite{faus,chem,new1,new2}. 

A special class of electronic excitations at metal surfaces are the so-called
image-potential states, which are localized on the vacuum side of the crystal
but can move almost freely in a plane parallel to the surface
\cite{echenique,smith}. These surface states occur when an electron is
trapped between its own image potential on the vacuum side and
a projected band gap of available bulk electronic states on the crystal side
of the surface. Because of their well-defined physical properties, image states
are useful to investigate the coupling of excited surface
electronic states with the substrate in the time domain \cite{plummer,petek}.

Recently, femtosecond ultraviolet-pump and infrared-probe techniques were
combined with two-photon photoemission (2PPE) to provide accurate measurements
of the relaxation dynamics of image-potential states on the (100) surfaces of
Cu and Ag  \cite{hofer1,hofer2}; the lifetime of the $n=1$ image state on these
surfaces was found to be $40\pm 6$ and $55\pm 5$ fs, respectively,
corresponding to linewidths of approximately 16.5 and 12 meV. Recent
theoretical work based on many-body calculations of the electron self-energy,
which revealed the main factors that determine the coupling of image states
to the substrate, predicted image-state lifetimes that are in the case of Cu
\cite{imanol1,imanol2,fauster1} and other metal surfaces
\cite{fauster2,fauster3,german} in good agreement with experiment but failed to
explain the measured relaxation of image-state electrons in Ag \cite{german}. 

In this Letter, we investigate the decay of the Ag(100) $n=1$ image state by
treating Auger-like single-particle and collective surface excitations on the 
same footing. A consistent treatment of these decay channels is particularly
important because of the near-degeneracy of the image state energy with the 
Ag(100) surface plasmon. The surprising and novel result of our work is that, 
although the imaginary part of the image-state self-energy is enhancend due 
to the presence of the plasmon decay channel, interferences resulting from 
the non-locality of the self-energy lead to a smaller overall image-state 
broadening, in agreement with experiment.      

While measured image-state lifetimes are found to be systematically longer 
in Ag than in Cu surfaces \cite{hofer2}, previous theoretical
estimates seemed to indicate that image-state lifetimes should be shorter
in the case of Ag. Although image states on Ag surfaces have energies
that are slightly closer to the Fermi level than in the case of Cu, both a
smaller density of $sp$ valence electrons and a smaller polarizability of $d$
bound electrons in Ag are expected to yield weaker screening and therefore
shorter image-state lifetimes in this material, as occurs in the case of bulk
excitations \cite{ekardt,idoia}. Furthermore, image-state lifetimes in Ag were
expected to be further reduced, due to the opening of the surface-plasmon
excitation channel\cite{liebsch}. As we argue below, this anticipated trend 
is reversed once the crucial nonlocality of the self-energy in the surface 
region is taken into account.  

A first-principles description of electron dynamics in the bulk of noble metals
shows that deviations from electron dynamics in a gas of $sp$ electrons mainly
originate in the participation of $d$ electrons in the screening of
electron-electron interactions \cite{campillo1,campillo2}; hence, we
account for the presence of occupied $d$ bands in Ag via a polarizable medium
giving rise to additional screening, as suggested in Ref.~\cite{liebsch0}, 
and compute the lifetime from the quasiparticle self-energy. 

The inelastic lifetime of image-state electrons at a solid surface is known to
be determined by electron-electron many-body interactions.  On the {\it
energy-shell} (i.e., neglecting the quasiparticle energy renormalization) and
assuming translational invariance in the plane of the surface, which is taken
to be normal to the $z$ axis, the decay rate (or lifetime broadening) of the
excited state $\phi(z)\,e^{i\,{\bf k_\parallel}\cdot{\bf r_\parallel}}$ with
parallel momentum ${\bf k_\parallel}$ and energy
$E=\varepsilon+k_\parallel^2/(2m)$ (unless stated otherwise, we use atomic
units, i.e., $e^2=\hbar=m_e=1$) is given by \cite{chem}
\begin{equation}\label{taum1}
\tau_{{\bf k}_\parallel,\varepsilon}^{-1}=-2\,\int dz\int
dz'\,\phi^{*}(z)\,{\rm Im}\Sigma_{{\bf
k}_\parallel,\varepsilon}(z,z')\,\phi(z'),  
\end{equation} 
where $\Sigma_{{\bf k}_\parallel,\varepsilon}(z,z')$ is the
two-dimensional Fourier  transform of the quasiparticle self-energy and $m$
represents the effective mass accounting for the potential variation in the
plane parallel to the surface. 

In the $GW$ approximation \cite{hedin} one considers only the first-order term
in a series expansion of the self-energy $\Sigma$ in terms of the screened
Coulomb interaction $W$. If one further replaces the Green function $G$ by that
of noninteracting electrons one finds  
\begin{equation}\label{selfener}  
{\rm Im}\Sigma_{{\bf k}_\parallel,\varepsilon}(z,z')=
\sum_{{\bf q}_\parallel,\varepsilon_f}\phi^{*}_f(z')\,{\rm
Im}W_{{\bf q}_\parallel,\omega}(z,z')\,\phi_f(z),
\end{equation}
where $\phi_f(z)\,e^{i\,{\bf k_\parallel}\cdot{\bf r_\parallel}}$ represent 
the available single-particle states with energy
$E_f=\varepsilon_f+({\bf k}_\parallel-{\bf q}_\parallel)^2/(2\,m_f)$ ($E_F\leq
E_f\leq E$), $\omega=E-E_f$ is the energy transfer, and $E_F$ is the Fermi
energy. Short-range exchange-correlation effects, which are absent in the GW
self-energy of Eq. (\ref{selfener}), can be included in the framework of the
so-called $GW\Gamma$ approximation \cite{mahan}. $GW\Gamma$ calculations for
the $n=1$ image-state lifetime on the (100) and (111) surfaces of Cu were
reported in Ref.~\cite{imanol2}, showing that they are very close to $GW$
calculations as long as the screened interaction $W$ is treated within the
random-phase approximation (RPA).

Here we report $GW$-RPA calculations of the electron self-energy and the
lifetime broadening by combining the one-dimensional model potential used in
previous work \cite{imanol1,imanol2}, which accurately accounts for the
dynamics of $sp$ valence electrons \cite{chulkov}, with the model
reported in Ref.~\cite{liebsch} in which the occupied $d$ bands are accounted
for by the presence of a polarizable medium which extends up to a certain
distance from the surface. Within this model, the screened interaction is of
the form $W=v'/(1-\chi^0v')$, where $v'$ is the $d$-screened Coulomb 
interaction and $\chi^0$ is the non-local dynamical susceptibility of 
noninteracting $sp$ electrons. We emphasize that this approach provides a
coherent treatment of single-particle and collective surface excitations.

The single-particle wave functions entering Eqs.~(\ref{taum1}) and
(\ref{selfener}), and those entering the expression for $\chi^0$, are
obtained by solving the one-dimensional Schr\"odinger equation modeled in 
Ref.~\cite{chulkov}. To derive $v'$ we assume that the $sp$ valence  electrons
are embedded in a polarizable background characterized by a $z$-dependent
local dielectric function $\epsilon_d(z,\omega)=\epsilon_d(\omega)$ for $z\leq
z_d$ and $\epsilon_d(z,\omega)=1$ for $z>z_d$, and find \cite{liebsch}    
\begin{eqnarray}\label{ll}
v'_{{\bf q}_\parallel,\omega}(z,z')&=&\frac{2\pi}
{q_\parallel\,\epsilon_d(z',\omega)}\,
[{\rm e}^{-q_\parallel\,|z-z'|}+{\rm sgn}(z_d-z')\nonumber\\
&\times&\sigma_d(\omega)\,{\rm e}^{-q_\parallel|z-z_d|}{\rm
e}^{-q_\parallel|z_d-z'|}], \end{eqnarray} where
$\sigma_d=[\epsilon_d(\omega)-1]/[\epsilon_d(\omega)+1]$.
$\epsilon_d(\omega)$ is related to the long-wavelength limit of the bulk
dielectric function via
$\epsilon(\omega)=\epsilon_{sp}(\omega)+\epsilon_d(\omega)-1$            
(see, e.g., Ref.~\cite{ehrenreich}), where $\epsilon_{sp}(\omega)$ corresponds
to the long-wavelength limit of the bulk dielectric function of
$sp$ valence electrons. The coordinate $z_d$ represents the
position of the boundary up to which the polarizable medium is assumed to
extend.

For comparison, we note that in a spatially uniform system the polarizable
medium extends to infinity ($z_d\to\infty$) and the $d$-screened Coulomb
interaction is simply $v'=v/\epsilon_d(\omega)$, $v$ being the bare Coulomb
interaction. The exact high-density limit of Eq.~(\ref{taum1}) is then found
to be, for electron energies near the Fermi level ($E-E_F<<E_F$),
$\tau^{-1}=\tau_{QF}^{-1}/\sqrt{\epsilon_d(\omega\to 0)}$, as suggested by
Quinn  \cite{quinn}, where $\tau_{QF}$ represents the formula first derived by
Quinn and Ferrel for a free-electron gas \cite{qf}.

The $n=1$ image state on Ag(100) is located close to the center of the
projected band gap, it has a binding energy of 0.53 eV, and
$\varepsilon-E_F=3.9\,{\rm eV}$. Its
probability-density has a maximum at $3.8\,$\AA\ outside the crystal edge
($z=0$), which we choose to be located half a lattice spacing beyond the last
atomic layer, and the penetration into the bulk crystal ($z<0$) is found to be
$5\%$, as in the case of Cu(100). The effective mass of the $n=1$ image
state on Ag(100) is known to be very close to the free-electron mass ($m \sim
1$), and for all single-particle energies entering $\chi^0$
and ${\rm Im}[-\Sigma]$ we choose the effective mass $m_f$ to increase from our
computed value of 0.45 at the bottom of the gap to the free-electron mass at
the bottom of the valence band. 
The local dielectric function $\epsilon_d(\omega)$ is taken from bulk optical
data \cite{johnson}. The boundary of the polarizable medium is taken to be at
$z_d=-1.5\,a_0$ ($a_0$ is the Bohr radius) inside the crystal, which was
previously found to best reproduce the anomalous dispersion of Ag surface
plasmons \cite{liebsch0}.

The lifetime broadening of the $n=1$ image state at the $\bar\Gamma$ point on
Ag(100) originates in processes in which the image-state electron with
energy $E=E_F+3.9\,{\rm eV}$ decays into an empty state with energy
$E_f$ above the Fermi level ($E_F\leq E_f\leq E$). These processes can be
realized by transfering energy and momentum to an excitation of the medium,
thereby creating either a single-particle or a collective surface excitation
with energy $\omega=E-E_f$ ($0\leq\omega\leq 3.9\,{\rm eV}$). Without
$d$-electron screening the surface-plasmon energy is too large
($\omega_{sp}=6.5\,{\rm eV}$) and only single-particle decay takes place; the
presence of $d$ electrons reduces $\omega_{sp}$ to $\sim 3.7\,{\rm eV}$ so
that relaxation via surface plasmons becomes feasible.

\begin{figure}
\includegraphics[width=0.45\textwidth,height=0.3375\textwidth]{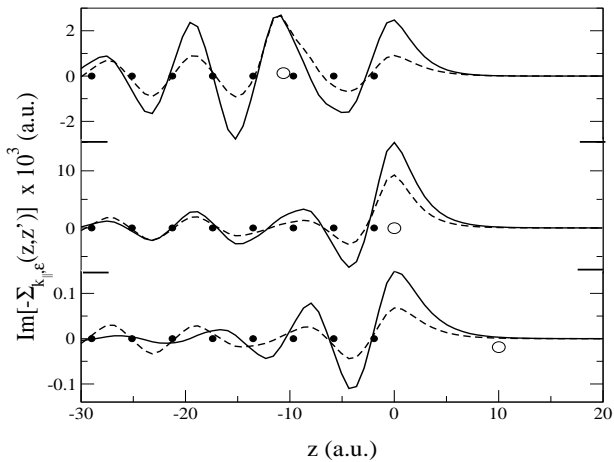}
\caption{Imaginary part of the $n=1$ image-state self-energy ${\rm
Im}[-\Sigma_{{\bf k}_\parallel,\varepsilon}(z,z')]$, versus
$z$, in the vicinity of the (100) surface of Ag . The value of $z'$ (indicated
by an open circle) is fixed at -10 (top panel), 0 (middle panel), and 10 $a_0$
(bottom panel). Notice the different scales.
Solid and dashed lines represent the results obtained with
$z_d=-1.5\,a_0$ and with $z_d\to -\infty$, respectively. 
Full circles represent the atomic positions. 
The geometrical electronic edge ($z=0$) is chosen to be located
half an interlayer spacing beyond the last atomic layer. 
Parallel momentum and energy of the quasiparticle are taken to be 
${\bf k}_\parallel=0$ and $\varepsilon=E_F+3.9\,{\rm eV}$.}
\label{fig2}
\end{figure} 

Fig.~\ref{fig2} shows the imaginary part of the self-energy 
${\rm Im}[-\Sigma_{{\bf k}_\parallel,\varepsilon}(z,z')]$ for ${\bf
k}_\parallel=0$ and $\varepsilon-E_F=3.9\,{\rm eV}$, corresponding to the
$n=1$ image state at the $\bar\Gamma$ point of the Ag(100) surface, as a
function of $z$ and for fixed values of $z'$. In the top panel
$z'$ is taken to be well inside the metal ($z'=-10\,a_0$), showing that ${\rm
Im}[-\Sigma]$ has a maximum at $z=z'$, as in the case of a homogeneous
electron gas. When $z'$ is fixed at the crystal edge ($z'=0$), as shown in the
middle panel of Fig.~\ref{fig2}, ${\rm Im}[-\Sigma]$ is still found to be
maximum at $z=z'$, but the magnitude of this maximum now being enhanced
because of the reduced electronic screening near the surface. The bottom panel
of Fig.~\ref{fig2} corresponds to $z'$ outside the metal
($z'=10\,a_0$). In this case, the main peak of ${\rm Im}[-\Sigma]$ lags
behind and remains localized at $z\sim0$ rather than $z=z'$, showing a
highly non-local behaviour of the imaginary part of the self-energy in the
presence of a metal surface. The largest values of ${\rm Im}[-\Sigma]$ are
obtained for $z,z'\approx 0$, because of the high probability of
single-particle and  collective electronic excitations in the surface region.

The imaginary part of the image-state self-energy in the absence of $d$ electrons
is also represented in Fig.~\ref{fig2}, as obtained with $z_d\to -\infty$ 
[or, equivalently, $\epsilon_d(\omega)=1$]. Although $d$ electrons
giving rise to additional screening should reduce both $W$ and ${\rm
Im}[-\Sigma]$, this reduction is outweighted by the opening of the 
surface-plasmon excitation channel which yields an overall enhancement of
${\rm Im}[-\Sigma]$. 

Within a local picture of the self-energy and assuming, therefore, that
${\rm Im}[-\Sigma]$ near $z\approx z'$ dominates the spectrum, it follows 
from Eq.~(\ref{taum1}) that an overall enhancement of the self-energy
yields an accordingly larger expectation value of ${\rm Im}[-\Sigma]$,  
i.e., an enhanced lifetime broadening. The key point of our work is, 
however, that this trend is reversed, due to the characteristic non-locality 
of the self-energy near the surface. As indicated in Table I, the calculated
lifetime broadening is 12\,meV, while without $sd$ screening and without 
plasmon decay the broadening is much larger, namely, 18\,meV.

\begin{table}
\caption{Calculated lifetime broadening $\hbar\,\tau^{-1}$ in meV, of 
the $n=1$ image state on Ag(100) with either $\epsilon_d(\omega)=1$ (or,
equivalently, $z_d\to -\infty$) or $\epsilon_d(\omega)$ from Ref.
\cite{johnson} and $z_d=-1.5\,a_0$. Contributions to the linewidth from
excitations where $0\leq\omega\leq 3.5\,{\rm eV}$ are displayed in parentheses.
The experimental linewidth is taken from time-resolved 2PPE experiments
\cite{hofer2}. The lifetime in fs ($1\,{\rm fs}=10^{-15}\,{\rm s}$) is
obtained by noting that $\hbar=658\,{\rm meV}\,{\rm fs}$.}
\begin{ruledtabular} \begin{tabular}{cccccc} $z_d$&local part &non-local
part&total&experiment\\  \hline $-\infty$ & 34(25) & $-16(-10)$ & 18(15) &  
\\ $-1.5$ & 59(25) & $-47(-15)$ & 12(10) & 12\\ \end{tabular}
\end{ruledtabular}
\label{table1}
\end{table}

To understand this puzzling result we note that according to Fig.~1 as 
long as $z'<0$ the self-energy exhibits local character. 
For $z'>0$, however, non-local behavior dominates. To distinguish these two 
spatial regions,  we refer to the broadening contribution arising from $z,z'<0$ 
as the 'local' part and the remainder ($z>0\ or\ z'>0$) as the 'non-local' part. 
These contributions to the lifetime broadening are also given in Table I. 
We point out that non-local effects give rise to interference contributions 
to the lifetime broadening, which are determined by the specific shape of the 
image-state wave function in the surface region where ${\rm Im}[-\Sigma]$ is 
maximum.

\begin{figure}
\includegraphics[width=0.45\textwidth,height=0.3375\textwidth]{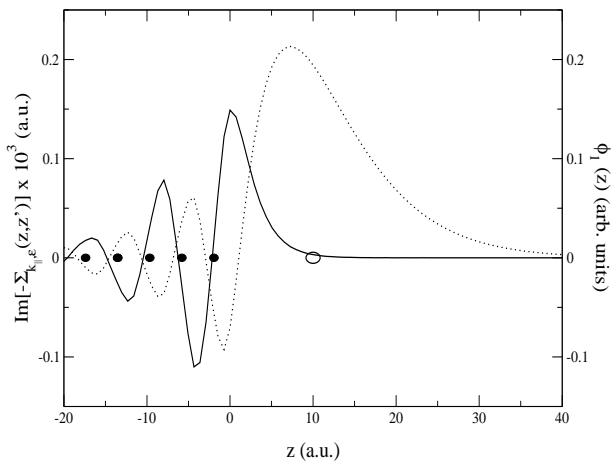}
\caption{Imaginary part of the $n=1$ image-state self-energy ${\rm
Im}[-\Sigma_{{\bf k}_\parallel,\varepsilon}(z,z')]$, versus $z$, in the vicinity of the
(100) surface of Ag, for $z'=10\,a_0$ and $z_d=-1.5\,a_0$. The dotted line
represents the $n=1$ image-state wave function. As in Fig.~\ref{fig2}, full
circles represent the atomic positions and the value of $z'$ is indicated by
an open circle. ${\bf k}_\parallel=0$ and
$\varepsilon=E_F+3.9\,{\rm eV}$.} 
\label{fig3}
\end{figure} 

This can be seen more clearly in Fig.~\ref{fig3} which shows the imaginary 
part of our calculated $n=1$ image-state self-energy on Ag(100) (see also 
the bottom panel of Fig.~\ref{fig2}), together with the $n=1$ image-state 
wave function. Since this function has a node at $z \sim 1\,a_0$ the major 
interference contributions to the lifetime broadening coming from $z'>0$ 
and $z\neq z'$ must be negative. Furthermore, as interference 
dominates the image-state relaxation when $z>0$ or $z'>0$, the lifetime 
broadening is actually smaller than one would have expected from the
penetration of the image state. According to the results given in Table I,
the role that $d$ electrons play enhancing the magnitude of the image-state 
self-energy is more pronounced on the vacuum side of the surface (where 
negative interference dominates) than in the bulk. The net effect of 
decay via surface plasmons and non-locality of the self-energy is therefore 
a considerably reduced lifetime broadening, in excellent agreement with the
time-resolved 2PPE measurement reported in Ref.~\cite{hofer2}.

To investigate in more detail the impact of surface-plasmon excitation on the
image-state lifetime broadening, we also give in Table \ref{table1} the 
calculated linewidths obtained only from low-energy excitations ($0\leq\omega\leq
3.5\,{\rm eV}$) where the surface-plasmon relaxation channel is excluded
(numbers in parentheses). In this case, the impact of $d$ electrons on the 
lifetime broadening is much smaller. Thus, 
the opening of the surface-plasmon relaxation channel, which only occurs when
both $d$ electrons and all contributions with $0\leq\omega\leq 3.9\,{\rm eV}$
are included, yields a considerable enhancement of both the 'local' and the
'non-local' part of the linewidth. As the large enhancement of the negative
'non-local' linewidth outweighs the increase of the 'local' contribution, the
overall influence of the occupied $d$ bands results in a reduced lifetime
broadening.

In conclusion, we have investigated the influence of single-particle
and collective surface excitations on the decay of image-potential states on
silver surfaces, by combining an accurate description of the dynamics of $sp$
valence electrons with a physically motivated model in which the occupied $d$
bands are accounted for by the presence of a polarizable medium. Our results
demonstrate that despite the enhancement of the image-state self-energy due to
decay via surface plasmons, the highly nonlocal character of this self-energy
in  the surface region ultimately leads to a surprisingly small lifetime
broadening, in agreement with the experimental data.

We acknowledge partial support by the University of the Basque Country, the
Basque Unibertsitate eta Ikerketa Saila, the Spanish Ministerio de Educaci\'on
y Cultura, and the Max Planck Research Award Funds. This work is dedicated to
Prof. Domingo Gonz\'alez from the University of Zaragoza on the ocassion of
his retirement.

\end{document}